# Hybrid Quantum-Classical Optimization of the Resource Scheduling Problem

Tyler Christeson[a], Md Habib Ullah[b], Ali Arabnya[c], Amin Khodaei[a], Rui Fan[a,1,]

[a]*Department of Electrical and Computer Engineering, University of Denver, Denver, 80210, CO, USA*
[b]*Department of Electrical Engineering and Electrical Engineering Technology, Penn State Harrisburg, Middletown, 17057, PA, USA*
[c]*Quanta Technology, Raleigh, 27607, NC, USA*

**Abstract**

Resource scheduling is critical in many industries, especially in power systems where the Unit Commitment (UC) problem determines the on/off status and output levels of generators under physical and economic constraints. Traditional exact methods, such as Branch-and-Bound, Branch-and-Cut, dynamic programming and mixed-integer linear programming (MILP), remain the backbone of UC solution techniques, but they often rely on linear approximations or exhaustive search, leading to high computational burdens as system size grows. Metaheuristic approaches, such as genetic algorithms, particle swarm optimization, and other evolutionary methods, have been explored to mitigate this complexity; however, they typically lack optimality guarantees, exhibit sensitivity to initial conditions, and can become prohibitively time-consuming for large-scale systems. In this paper, we introduce a quantum-classical hybrid algorithm for UC—and, by extension, other resource scheduling problems—that leverages Benders decomposition to decouple binary commitment decisions from continuous economic dispatch. The binary "master" problem is formulated as a quadratic unconstrained binary optimization (QUBO) model and solved on a quantum annealer. The continuous "subproblem," which minimizes generation costs, with Lagrangian cuts feeding back to the master until convergence. We evaluate our hybrid framework on systems scaled from 10 to 1,000 generation units. Compared against a classical mixed-integer nonlinear programming (MINLP) baseline, the hybrid algorithm achieves a consistently lower computation-time growth rate and maintains an absolute optimality gap below 1.63%. These results demonstrate that integrating quantum annealing within a hybrid quantum-classical Benders decomposition loop can significantly accelerate large-scale resource scheduling without sacrificing solution quality, pointing toward a viable path for addressing the escalating complexity of modern power grids.

## 1. Introduction

Resource scheduling refers to a broad class of problems that involves the strategic allocation of limited resources—whether they be time, materials, personnel, or energy—to meet a set of forecasted activities or demands [1]. This problem class is present in diverse fields including industrial manufacturing [2], construction planning [3], cloud computing management [4], and, notably, power systems operations [5]. In the realm of power systems, the ongoing transformation driven by the increasing proliferation of distributed energy re- sources (DERs) such as smart grids [6] and electric vehicles [7, 8], the shift toward cleaner technologies [9], and the need for enhanced resilience against extreme weather events [10] has underscored the need for updated management and decision-making practices.

*Email address:* rui.fan@du.edu (Rui Fan)

Among the various resource scheduling challenges within power systems, unit commitment (UC) stands out as a critical decision-making process. UC involves the optimal scheduling of generating units over different time intervals in order to meet forecasted loads while adhering to physical, environmental, and economic constraints. The UC problem embodies the complexity inherent in resource scheduling; it is proven to be NP-hard [11], and its combinatorial structure leads to significant computational challenges, especially as systems scale and incorporate higher levels of renewable integration.

This computational complexity has profound economic implications. Even modest improvements in the accuracy and efficiency of UC solutions can yield substantial cost savings for system operators and, consequently, end consumers. Mixed-integer linear programming (MILP) remains one of the most commonly used techniques for solving the UC [12, 13]. However, most MILP models for UC problems are developed by approximating a unit's nonlinear production cost as a piece-wise linear function; the optimal solutions of these approximated MILPs are never the optimal solutions to the original UC problem, even for small systems [14]. Although the Branch-and-Bound [15] and Branch-and-Cut [16] methods can be adopted to find optimal UC solutions, these techniques are often time-consuming for large-scale systems and may encounter computational efficiency barriers [17, 18]. The UC problems have also been solved using dynamic programming that necessitates continuously computing the value of the same subproblems in order to reach the optimum solution [19]. However, the "curse of dimensionality" increasingly becomes a challenge as the size of the system and the number of decision variables increases [20]. The evolution-inspired search methods such as Genetic Algorithm [21], and Particle Swarm Optimization [22], among others [23], were adopted to improve the UC solution as well. These methods can find a satisfactory solution, but are time-consuming, do not guarantee optimality, and are sensitive to initial value selections [24].

While the existing UC solutions have increased their accuracy and computational efficiency, it is worth noting that they have not kept pace with the increasing complexity of this problem due to the ongoing transition of the modern power grids. That lag can be primarily attributed to the limitations of classical computing technologies. On the other hand, given the recent breakthroughs in quantum computing (QC) technology and their capabilities in solving combinatorial optimization problems, they can be considered as a viable alternative to address this issue [25, 26]. In particular, quantum annealing (QA) has shown potential in efficiently solving combinatorial optimization problems [27], such as UC, by exploring vast solution spaces through quantum mechanical phenomena. However, a QA solution has previously been tested for UC and found to be inaccurate and inefficient in computation time [28] due to having to discretize continuous variables when formulating problems as quadratic unconstrained binary optimization (QUBO) models for compatibility with current quantum hardware. Although QA alone may not fully capture the intricate constraints of UC, integrating it with classical computing approaches into a hybrid model can leverage the strengths of both paradigms. To accomplish this, the Benders decomposition method is used to decouple binary commitment decisions from continuous economic dispatch. Benders decomposition is a mathematical technique that partitions a complex mixed-integer problem into two interconnected subproblems: a "master" problem handling the complicating integer variables and a "subproblem" addressing the continuous variables [29]. The iterative nature of the method allows for efficient convergence by successively refining the feasible region of the master problem using optimality and feasibility cuts derived from the subproblem. This approach is particularly effective for unit commitment problems, where binary commitment decisions can be naturally separated from continuous power dispatch variables, and has been investigated as a method for hybrid quantum-classical algorithms, even for UC.

In [30], the authors demonstrate a hybrid quantum-classical algorithm with Benders' decomposition to solve a MILP problem, but do not apply it to a specific domain. Our work applies a similar framework to the UC problem, targeting large-scale optimization efficiency. Unlike [30], which discretizes continuous variables in the master problem, our approach separates binary and continuous variables into the master and subproblems, respectively. This reduces qubit requirements and improves scalability on current QPUs. In [31], the authors also apply a hybrid quantum-classical Benders decomposition to UC but encode continuous variables as binaries in the master problem. Our approach instead optimizes binary variables on the quantum master and continuous variables on the classical subproblem, enabling more efficient qubit usage and avoiding loss of precision from discretization. Furthermore, [31] tests only on a small system (two units, five time steps), whereas we evaluate up to 1,000 units over 24 hours, demonstrating scalability and potential real-world application. In [32], the authors propose hybrid algorithms for UC in distributed microgrids, incorporating coupling/decoupling with the utility grid. Our work instead develops a single hybrid algorithm for UC in a centralized transmission network, focusing on large-scale generation scheduling. While [32] successfully models the IEEE-RTS-24 system with 99 distributed resources, our framework scales to 1,000 generation units, moving closer to real-world system operations and highlighting the potential of quantum computing in large-scale power system optimization.

Importantly, our use of QA is motivated not solely by performance comparison with mature classical metaheuristics but by the opportunity to evaluate QA's role within a hybrid optimization workflow and explore its potential as quantum hardware continues to advance. QA is well-suited to the combinatorial structure of the UC master problem and, when combined with Benders decomposition, allows the continuous economic dispatch subproblem to remain on a classical solver without discretization. Additionally, this work leverages D-Wave's Constrained Quadratic Model (CQM) framework, which automates constraint weighting and mitigates the need for manual penalty tuning— an important practical consideration when scaling to large problem sizes. Although QA may not yet consistently outperform state-of-the-art classical approaches, this study positions QA as a forward-looking technology and assesses its ability to handle large-scale UC formulations as hardware capability improves.

We propose a quantum-classical hybrid method in which a Benders decomposition technique is used to separate the binary variables and continuous variables so that the problems associated with respective variables can be solved in a quantum annealer and in a classical computer, accordingly. To address the gap in existing quantum annealing UC solutions, this paper develops a novel framework for quantum-compatible UC solutions and investigates a set of case studies to further investigate this issue. The major contributions of this paper can be summarized as follows:

- We propose a new quantum-compatible hybrid framework for solving the UC problem, which utilizes both quantum and classical computing approaches to achieve a faster computation time than classical-only solutions and greater accuracy than quantum-only solutions at larger system sizes.
- With the new hybrid framework, we successfully simulated system sizes up to 1,000 generation units, marking an improvement over previous QCUC methods and a step closer to real-world implementation for QCUC.
- We established and streamlined a systematic and efficient process of transforming linear/nonlinear programming models for large-scale resource scheduling problems, like UC, into QUBO and hybrid models that can be used for quantum applications.

The remainder of this paper is organized as follows. Section II focuses on unit commitment as a representative case of resource scheduling, discussing its formulation and computational challenges. Section III introduces the basics of quantum computing and quantum annealing, their application to combinatorial optimization within power systems, and discussions of methods for converting classical formulations into quantum-compatible forms. Section IV then presents the Benders decomposition method for our proposed quantum-compatible hybrid framework for UC, detailing the transformation of classical models into a hybrid form amenable to quantum processing. Section V analyzes the performance of the proposed model against traditional methods, and finally, Section VI concludes with insights on future research directions aimed at further advancing computational approaches in resource scheduling.

## 2. Unit Commitment Formulation

The UC problem is a form of resource scheduling problem, with the objective of minimizing the total cost of operating a set of power generation units over a given timeframe while subject to a variety of constraints. These constraints consist of economic, environmental, operational, and physical constraints. The mathematical formulation of UC problems as given by [33, 34, 35, 36] can be seen represented here:

$$\min_{p_{it}, v_{it}} \sum_{i \in \mathcal{N}} \sum_{t \in \mathcal{T}} [C_i(p_{it}) + U_i(p_{it}) + D_i(p_{it})] \tag{1a}$$

s.t.:

$$\sum_{i \in \mathcal{N}} p_{it} = L_t, \quad \forall t \tag{1b}$$

$$P_i^{min} v_{it} \leq p_{it} \leq P_i^{max} v_{it}, \quad \forall i, \quad \forall t \tag{1c}$$

$$\sum_{i \in N} P_i^{max} v_{it} \geq L_t + S_t, \quad \forall t \tag{1d}$$

$$-R_i^{dn} \leq p_{it} - p_{it-1} \leq R_i^{up}, \quad \forall i \tag{1e}$$

$$U_i^{\tau} \leq \tau_i^{up}, \quad \forall i \tag{1f}$$

$$D_i^{\tau} \leq \tau_i^{dn}, \quad \forall i \tag{1g}$$

$$F_i^{min} \leq \sum_{t \in \mathcal{T}} \Xi_i(p_{it}) + \Lambda_{it} \leq F_i^{max}, \quad \forall i \tag{1h}$$

$$\Pi_i^{min} \leq \sum_{i \in \mathcal{N}} \sum_{t \in \mathcal{T}} \Xi_i(p_{it}) + \Lambda_{it} \leq \Pi_i^{max} \tag{1i}$$

$$\sum_{i \in \mathcal{N}} \sum_{t \in \mathcal{T}} \Lambda_i(p_{it}) + \Psi_{it} \leq E^{max} \tag{1j}$$

where (1a) gives the objective function representing the overall costs of the system consisting of the power generation, start-up, and shut-down costs denoted as $C_i(p_{it})$, $U_i(p_{it})$, and $D_i(p_{it})$, respectively, for the generating unit $i \in N$. The power generation cost $C_i(p_{it})$ is given by a quadratic function as: $C_i(p_{it}) = a_i v_{it} + b_i p_{it} + c_i p_{it}^2$. The start-up and shut-down costs ($U_i(p_{it})$ and $D_i(p_{it})$) can be designed as described in [34].

Eq. (1b) gives the power balance constraints of the system, and (1c) is the minimum and maximum power limits of the generating units. Eqs. (1d) and (1e) represent the spinning reserve of the system and generators' ramping rate constraints, respectively. The generators are constrained by minimum uptime and downtime constraints, which are expressed by (1f) and (1g), respectively. The fuel limit constraints of the entire system, as well as each generating unit, are given by (1h) and (1i), respectively. The constraint (1j) expresses the total emission limit of the given system. Further details of these constraints can be found in [34]. There can be a set of other constraints incorporated in the UC problem, including but not limited to transmission line capacity for real and reactive power flows, bus voltage amplitude and phase limits, transmission network maintenance constraints, and load shedding [33, 34, 35, 36, 37].

## 3. Quantum Annealing (QA) and QA-Compatibility

*3.1 Quantum Annealing*

Quantum annealing (QA) is a heuristic approach to solving optimization problems based on adiabatic quantum computing. It takes advantage of quantum physics phenomena such as quantum tunneling, entanglement, and superposition. The QA finds the global minimum of a given objective function over a range of solutions, and is mostly utilized for problems with a discrete search space (combinatorial optimization problems) and a large number of local minima [38]. The QA can be designed for binary quadratic model (BQM) (to minimize QUBO or Ising functions), and for discrete quadratic model (DQM) as [39, 40]:

BQM (QUBO):
$$min \sum_{ij} Q_{ij} x_i x_j + \sum_i m_i x_i, \tag{2a}$$

BQM (Ising):
$$min \sum_{ij} J_{ij} y_i y_j + \sum_i h_i y_i, \tag{2b}$$

DQM:
$$min \sum_{ij} G_{ij}(d_i d_j) + \sum_i n_i(d_i), \tag{2c}$$

where $x_i \in \{0,1\}$ and $y_i \in \{-1,1\}$ are the decision variables, $m_i$ and $h_i$ are the linear weight, and $J_{ij}$ and $Q_{ij}$ for $i, j \in \{1, 2, ..., n\}$ is the quadratic coupler specified by the users to define the considered problem. The QUBO and Ising formulations are equivalent by $x_i = (1 + y_i)/2$ [41]. In (2c), $d_i$ is the discrete variable, $n_i(.)$ and $G_{ij}(.)$ are real-valued functions. The DQM can be transformed to equivalent binary model by replacing the discrete variable $d_i$ with binary variable using one-hot encoding constraint $\sum_n x_{i(n)} = 1, \forall i$ which can further be formulated introducing penalty parameter [40, 42, 43].

$$min \sum_{ij} \varphi_{ij} \rho_i \rho_j + \sum_i \psi_i \rho_i \tag{3a}$$

$$\sum_I \alpha_{i(\varepsilon)} \rho_i + \sum_{ij} \beta_{ij(\varepsilon)} \rho_i \rho_j = 0, \ \varepsilon = 1, ..., \Phi_{eq} \tag{3b}$$

$$\sum_i \phi_{i(\epsilon)} \rho_i + \sum_{ij} \varsigma_{ij(\epsilon)} \rho_i \rho_j \leq 0, \ \epsilon = 1, ..., \Phi_{inq} \tag{3c}$$

where decision variable $\rho_i$ can be binary or real-valued integer variable, $\varphi_{ij}, \psi_i, \alpha_i, \beta_{ij}, \phi_{i(\epsilon)}, \varsigma_{ij(\epsilon)}$ are real-valued coefficients, and $\Phi_{eq}$ and $\Phi_{inq}$ are the number of equality and inequality constraints, respectively. The QA algorithm can be implemented as a time-varying Hamiltonian $\mathcal{H}$ with a $t_a$ time interval as [45]:

$$\mathcal{H}(s) = A(s) H_I + B(s) H_P \tag{4}$$

where ta is the annealing time during which $\mathcal{H}$ transitions from $H_I$ to $H_P$, annealing path functions $A(s)$ and $B(s)$ are defined in terms of normalized time $s = t/t_a$, which satisfy the conditions $A(0) = 1$, $A(t_a) = 0$ and $B(0) = 0$, $B(t_a) = 1$. The initial Hamiltonian $H_I$ describes initial conditions, which can be given by

$$H_I = \sum_i \sigma_x^{(i)} \tag{5}$$

where $\sigma_x^{(i)}$ is the Pauli-x operator applied to qubit $i$. The problem Hamiltonian $H_P$ describes the ground state for the qubits, which can be given as:

$$H_P = \sum_i h_i \sigma_z^{(i)} + \sum_{ij} J_{ij} \sigma_z^{(i)} \sigma_z^{(j)} \tag{6}$$

where $\sigma_z^{(i)}$ denotes the Pauli-z operator applied to qubit $i$. The vectors and eigenvalues of $H_P$ correspond to the solutions and costs defined by (2).

The QA process works as follows. The qubits are placed in a superposition state at $s = 0$ according to $H_I$. When $s$ increases, $\mathcal{H}(s)$ develops by increasing $B(s)$ and diminishing $A(s)$, which allows the problem Hamiltonian to eventually evolve. When $s = 1$ at the end, the qubit states give classical spin values $y_I = \{\pm 1\}$ and a minimum-eigenvalue (ground) state for $\mathcal{H}(s = 1) = H_P$ corresponds to an ideal solution for the energy function of the Ising model (2b) [46]. As quantum annealers are open systems, there is a chance that the qubits would not end in the ground state; hence in the common application, multiple anneals are used to improve the solution quality. To solve a problem using a QA approach on D-Wave's quantum annealers, the objective function should first be formulated as BQM, DQM, or CQM. Then the objective function's variables such as $x_i$ or $y_i$ or $\rho_i$ are mapped to qubits on the QPU, and the quadratic coefficients such as $Q_{ij}$ or $J_{ij}$ or $G_{ij}$ or $\varphi_{ij}$ are mapped to couplers, which is called minor embedding. Finally, the QPU utilizes QA to find the solution to the problem.

### 3.2 QA-Compatible Unit Commitment

In BQM, any optimization problem must be reformulated as a quadratic unconstrained binary optimization. Therefore, the UC problem in (6) is required to be transformed into a QUBO problem to solve it in a quantum annealer with BQM solvers. This paper presents two methods to design a quantum-compatible UC (QCUC) problem in the form of QUBO, described in the following sub-sections. The QUBO for the UC problem through the discretization process can be achieved using two different approaches, as follows. The QC formulation can be given as:

$$\min \sum_i \left(a_i(1 - v_i) + b_i p_i + c_i p_i^2\right) \tag{7}$$

s.t.:

$$p_i = \sum_{k=1}^{N+1} \left(p_i^{min} + h_i(k-1)\right) z_{ik} \quad \forall i \tag{8a}$$

$$h_i = \frac{p_i^{max} - p_i^{min}}{N} \quad \forall i \tag{8b}$$

$$\sum_i p_i = L \tag{8c}$$

$$v_i + \sum_{k=1}^{N+1} z_{ik} = 1 \quad \forall i \tag{8d}$$

$$\sum_t^{t+UT_i-1} (1 - v_{i,t}) \geq UT_i(v_{i,t-1} - v_{i,t}), \quad \forall i, \quad \forall t \in [1, \mathcal{T} - UT_i + 1] \tag{8e}$$

$$\sum_t^{t+DT_i-1} (v_{i,t}) \geq DT_i(v_{i,t} - v_{i,t-1}), \forall i, \forall t \in [1, \mathcal{T} - DT_i + 1] \tag{8f}$$

$$p_{i,t} - p_{i,t-1} \leq RU_i(1 - v_{i,t}), \forall i, \forall t \in [2, \mathcal{T}] \tag{8g}$$

$$p_{i,t-1} - p_{i,t} \leq RD_i(1 - v_{i,t-1}), \quad \forall i, \forall t \in [2, \mathcal{T}] \tag{8h}$$

where $a_i$, $b_i$, and $c_i$ are the binary, linear, and quadratic cost coefficients, respectively, and $v_{i,t}$ and $p_{i,t}$ are the binary and continuous decision variables representing the online status and power generation of unit $i$ at hour $t$. Eqs. (8a) and (8b) detail how the continuous variable $p_{i,t}$ is discretized, by splitting it into $N$ segments that range from $P^{min}$ to $P^{max}$ in increments of $h_i$. This introduces a new binary variable, $z_{i,t,k}$, for the online status of the $N$ discretized segments of the power generation of unit $i$ at hour $t$. Eq. (8c) represents the load constraint, analogous to (1b). Eq. (8d) constrains the binary variables $v_{i,t}$ and $z_{i,t,k}$ for each unit at each hour, ensuring that only one bin is selected. Eqs. (8e) and (8f) represent the minimum up-time and down-time constraints, analogous to (1f) and (1g). Eqs. (8g) and (8h) represent the ramping rate constraints, similar to (1e), although separated into two constraints to be reformulated into a QUBO model. Once the constraints are rewritten, they can be summed with the objective to form a single QUBO to be sent to the quantum annealer:

$$\begin{aligned}
\min_{p_{it}, v_{it}} \sum_{t \in \mathcal{T}} \sum_{i \in \mathcal{N}} &\left( a_i v_i + b_i \sum_{k=1}^{N+1} \left( P_i^{min} + h_i(k-1) \right) z_{i,t,k} + c_i \left( \sum_{k=1}^{N+1} \left( P_i^{min} + h_i(k-1) \right) z_{i,t,k} \right)^2 \right. \\
&+ A \left( \sum_i \sum_{k=1}^{N+1} \left( P_i^{min} + h_i(k-1) \right) z_{ik} - L \right)^2 + B \left( (1 - v_i) + \sum_{k=1}^{N+1} z_{ik} - 1 \right)^2 \\
&+ C \sum_{i \in \mathcal{N}} \left( UT_i(v_{i,t-1} - v_{i,t}) - \sum_{t}^{t+UT_i-1} (1 - v_{i,t}) \right)^2 \\
&+ D \sum_{i \in \mathcal{N}} \left( DT_i(v_{i,t} - v_{i,t-1}) - \sum_{t}^{t+DT_i-1} (v_{i,t}) \right)^2 \\
&+ E \sum_{i \in \mathcal{N}} \left( \sum_{k=1}^{N+1} (h_i(k-1)) z_{i,t,k} - \sum_{k=1}^{N+1} (h_i(k-1)) z_{i,t-1,k} - RU_i \right)^2 \\
&\left. + F \sum_{i \in \mathcal{N}} \left( \sum_{k=1}^{N+1} (h_i(k-1)) z_{i,t-1,k} - \sum_{k=1}^{N+1} (h_i(k-1)) z_{i,t,k} - RD_i \right)^2 \right)
\end{aligned} \tag{9}$$

where $A$-$F$ are the penalty coefficients associated with (8c)-(8h). These penalty coefficients ensure that their respective constraints are satisfied, and need to be weighted such that no individual penalty outweighs the others or the objective, leading to nonoptimal or infeasible solutions as shown in [47]. Eq. (9) is the QUBO formulation of the UC problem, with (8a), (8b) detailing the discretization used in this traditional QA-only QUBO approach.

### 3. Benders Decomposition

The proposed hybrid quantum-classical algorithm utilizes Benders decomposition, which is an iterative approach for solving large-scale optimization problems by dividing the overall optimization problem into several sub-problems, and introducing Lagrangian variables by simultaneously considering all decision variables and constraints [29]. One of the subproblems is called the master problem, and the solution to the master problem is sent to the subproblems to influence their solutions. Further subproblems send their solutions back to the master problem in the form of a cut, and information continues to be sent back and forth between master and subproblems in an iterative manner until the optimal unit commitment solution is converged upon. The formulation of the master problem for the UC problem presented in (7)-(8h) can be given as follows:

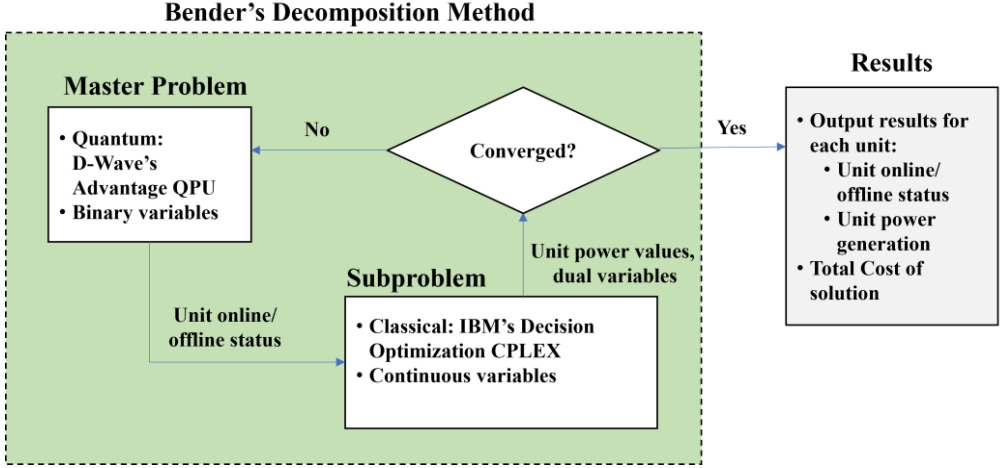

Figure 1: Process Flow for Quantum Decomposed Hybrid and MINLP solutions

$$\min_{v_{i,t}} w = \sum_{t \in \mathcal{T}} \sum_{i \in \mathcal{N}} (a_i v_{i,t}) + \lambda, \tag{10a}$$

$$\sum_{i \in \mathcal{N}} P_i^{max} v_{i,t} \geq L_t, \quad \forall t \tag{10b}$$

$$\sum_{t}^{t+UT_i-1} (v_{i,t}) \geq UT_i(v_{i,t} - v_{i,t-1}), \quad \forall i, \forall t \in [1, \mathcal{T} - UT_i + 1] \tag{10c}$$

$$\sum_{t}^{t+DT_i-1} (1 - v_{i,t}) \geq DT_i(v_{i,t-1} - v_{i,t}), \quad \forall i, \forall t \in [1, \mathcal{T} - DT_i + 1] \tag{10d}$$

where $\lambda \geq 0$ is a Lagrangian variable representing the solutions to the subproblem. The master subproblem determines the value of the binary variable $v_{i,t}$ based on minimizing the binary cost coefficient $a_i$, while ensuring that the units selected are able to satisfy each hour's load constraint with eq. (10b) and the minimum-required uptime and downtime of each unit with (10c) and (10d). Based on the values of $v_{i,t}$, indicating the selected units, the subproblem computes the value of $p_{i,t}$. The formulation for the subproblem can be given, as follows:

$$\min_{p_{i,t}} \sum_{t \in \mathcal{T}} \sum_{i \in \mathcal{N}} (b_i p_{i,t} + c_i p_{i,t}^2) \tag{11a}$$

$$u_{i,t} = \widehat{v_{i,t}} \quad \forall i, \forall t \tag{11b}$$

$$\sum_{\forall i \in \mathcal{N}} p_{i,t} = L_t \quad \forall t \tag{11c}$$

$$p_{i,t} - p_{i,t-1} \leq RU_i u_{i,t}, \quad \forall i, \forall t \in [2, \mathcal{T}] \tag{11d}$$

$$p_{i,t-1} - p_{i,t} \leq RD_i u_{i,t-1}, \forall i, \forall t \in [2, \mathcal{T}] \tag{11e}$$

where $u_{it}$ are the binary variables, now fixed as constants by taking their values from the initial master problem $\hat{v}_{it}$. Using the fixed selected units, the subproblem minimizes the economic dispatch of those units, while ensuring that the constraints around power generation, such as the power bound constraints, load satisfaction constraint, and ramping rate constraints, are satisfied. The solution from the subproblem is sent back to the master problem in the form of a cut. The cut is defined as:

$$\lambda \geq \hat{w} + \sum_t \sum_i \mu_{it}(v_{it} - \widehat{v_{it}}) \tag{12}$$

where $\hat{w}$ is the minimized objective value from the subproblem (11), $\mu_{i,t}$ is the dual variable value from each of the constraints for each unit $i$ at hour $t$, and $\hat{v}_{i,t}$ is the binary online status of unit $i$ according to the solution from the previous master problem. The cut is returned to a new iteration of the master problem, where units are newly selected using the information from the cut regarding dual variables from the constraints of previously chosen and optimized units. This iterative information exchange between the master and subproblem continue until the units selected in the master problem and the objective value from the economic dispatch in the subproblem remain constant throughout iterations. To utilize Benders decomposition in a hybrid quantum-classical model, the master problem of unit selection is chosen as the quantum component and must be converted into QUBO format. The final QUBO for the quantum master problem, is defined as:

$$\min_{v_{i,t}} \sum_{t \in T}\left(\sum_{i \in N}(a_i v_{i,t}) + A(\sum_{i \in N} P_i^{max} v_i - L) + B \sum_{i \in N}\left(UT_i(v_{i,t} - v_{i,t-1}) - \sum_{\tau=1}^{t+UT_i-1} v_{i,t}\right)^2 \right. \\ \left. + C \sum_{i \in N}\left(DT_i(v_{i,t-1} - v_{i,t}) - \sum_{\tau=1}^{t+DT_i-1}(1 - v_{i,t})\right)^2 + \lambda \tag{13}$$

where $\lambda$ is defined in the cut according to (12). The resulting QUBO for the master problem of the proposed method utilizes three penalty coefficients, unlike the QUBO for the QA-only method in (9) which utilizes six penalties. This will allow for easier balancing of penalty and objective terms in the decomposition hybrid model than in the previous discretization model. This would increase the likelihood of finding feasible solutions in methods like D-Wave's BQM that only utilize QUBO formulations. However, for this hybrid model, we utilize the CQM, which provides direct support for modeling mathematic constraints [44] and only returns solutions that are feasible according to those constraints. Because of this, it is not necessary to tune penalty coefficients like in previous work [47]. This capability to directly model constraints reduces the need for extensive manual tuning and improves workflow efficiency, while still maintaining feasible and high-quality solutions.

Additionally, since the continuous power generation variables are handled using classical linear programming solvers in the subproblem of this proposed method, the resulting power generation can be represented as precise solutions, rather than as approximations achieved by the discretized QA-only QUBO model. The result here is a quantum-classical hybrid method in which a Benders decomposition technique is used to separate the binary variables and continuous variables so that the problems associated with respective variables can be solved in a quantum annealer and in a classical computer, respectively. The iterative process flow for the hybrid model is visualized in Figure 1.

## 4. Numerical Results

The simulations in this paper are all based on a 10-unit system, and the parameters for this system can be seen in Table 1. The quantum annealing aspects were programmed using D-Wave's Ocean SDK and executed on the Advantage 4.1 QPU and its Pegasus topology [48]. The hybrid algorithm was formulated within D-Wave's CQM framework, and once variables were encoded in QUBO form, the model was solved using the LeapHybridCQMSolver with default parameters [49]. The Advantage 4.1 QPU has 5,640 qubits, which allowed us to model up to 1,000 units in a single system. This work intentionally benchmarks a 1,000-unit UC system as a practical upper bound for current quantum annealing hardware. Within the proposed Benders decomposition framework, each generating unit contributes three binary decision variables (online status, startup, and shutdown), yielding 3,000 binary variables for the largest test case—well within the 5,000-variable capacity supported by D-Wave's CQM [44]. While this fits comfortably within current hardware limits, we note that minor-embedding requires mapping each logical variable to multiple physical qubits, introducing connectivity overhead and potentially increasing embedding time. The Pegasus topology supports 16 couplers per qubit [48], meaning dense, highly connected problems require careful embedding to avoid qubit fragmentation.

Table 1: 10-Unit System Parameters

| $a$ [\$] | $b$ [\$/MWh] | $c$ [\$/MW²h] | $P^{max}$ [MW] | $P^{min}$ [MW] | $U_i^+$ [hr] | $D_i^-$ [hr] | $R_i$ [MW] |
|---|---|---|---|---|---|---|---|
| 1000 | 16.19 | 0.00048 | 455 | 150 | 8 | 8 | 150 |
| 970 | 17.26 | 0.00031 | 455 | 150 | 8 | 8 | 150 |
| 700 | 16.69 | 0.002 | 130 | 20 | 5 | 5 | 40 |
| 680 | 16.5 | 0.00211 | 130 | 20 | 5 | 5 | 40 |
| 450 | 19.7 | 0.00398 | 162 | 25 | 6 | 6 | 45 |
| 370 | 22.26 | 0.00712 | 80 | 20 | 3 | 3 | 20 |
| 480 | 27.74 | 0.0079 | 85 | 25 | 3 | 3 | 25 |
| 660 | 25.92 | 0.00413 | 55 | 10 | 1 | 1 | 15 |
| 665 | 27.27 | 0.00222 | 55 | 10 | 1 | 1 | 15 |
| 670 | 27.79 | 0.00173 | 55 | 10 | 1 | 1 | 15 |

The MINLP simulations utilized IBM's DOcplex. A genetic algorithm (GA) solution [50] and a simulated annealing (SA) algorithm [51] solution were simulated in order to provide comparisons to classical metaheuristic methods relevant to UC. All simulations were done using Python v3.10. The smallest simulation size analyzed is the 10-unit system presented in Table 1, and the largest simulation size analyzed is a 1,000-unit system. The simulations are scaled up in 10-unit increments by duplicating the original 10-unit system to reach the desired system size. This duplication process means that many units in scaled-up systems will be exactly the same and that larger systems will likely have more than one unit commitment schedule that achieves the minimum cost. Because of this, to compare the accuracy of the hybrid method, absolute relative cost error is used in terms of percentages, with the MINLP classical-only solution taken as the baseline. This allows multiple unit commitment schedules to be compared on accuracy to the optimal schedule using a single metric. For the MINLP and heuristic algorithm solutions, 10 runs were conducted at each system size and the solution time and final objective value were taken for each run. For the hybrid quantum-classical solution, 10 runs were conducted in the range of 10-unit to 100-unit systems, to characterize the variance and consistency of the hybrid algorithm. Due to limited access time on the CQM, we were unable to perform the 10 runs for larger systems, and systems larger than 100 units were restricted to a single run. For each run of the hybrid quantum-classical solution, computation time and objective value were taken for each iteration of the hybrid algorithm. Since the proposed hybrid quantum-classical algorithm is an iterative algorithm, it is necessary to show the progress of solution quality over several iterations and characterize the convergence behavior of the algorithm. The iterative process of the proposed hybrid model requires two consecutive iterations to be within 0.5% of the objective value of previous iteration's objective value in order to converge. Figure 2 shows the objective value per iteration count for system sizes ranging from 10 to 100 units, for 10 runs each. The difference in objective value is represented as the absolute value of the difference in an iteration's objective value from the previous iteration. For smaller system sizes (10-50 units), 3 iterations are often all that is needed to converge, and as the system size increases, more iterations are needed, up to 8 iterations in the maximum case. Generally, solutions converge upon an area within the iterative process, which is seen as the optimality gap tends to decrease with iterations.

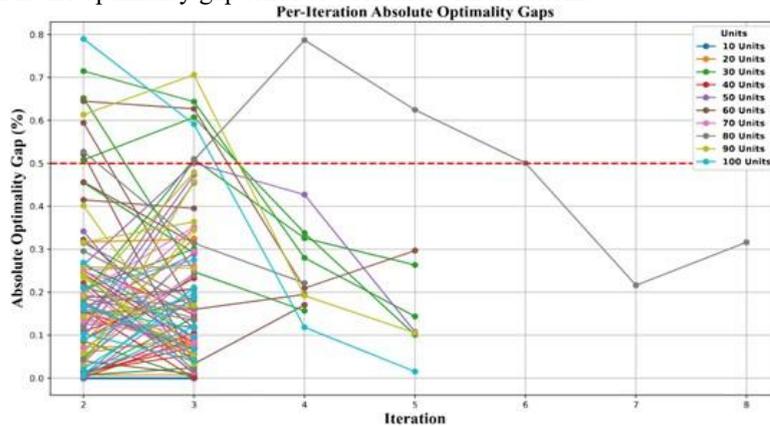

Figure 2: Per-Iteration Absolute Optimality Gap

A 24-hour cycle for load demand is considered, which can be seen in Figure 3. The load demand is also linearly scaled for any simulations of systems beyond 10 units.

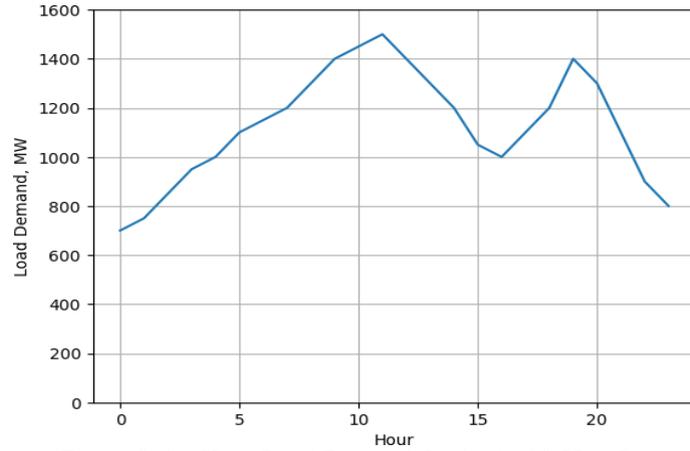
Figure 3: 24-Hour Load Demand Cycle for 10-Unit System

Figure 4 presents the computation times for the MINLP, GA, and SA classical methods and the proposed quantum–classical hybrid approach across a range of system sizes.

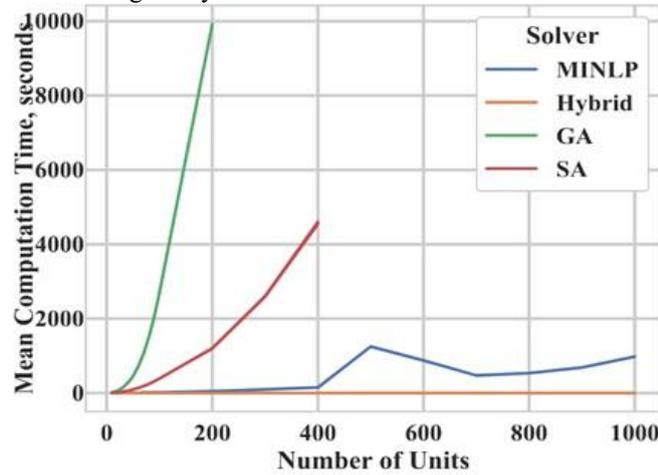
Figure 4: Computation Time of UC Solutions

For small- to medium-scale systems, ranging from 10 to 100 generation units, the MINLP and hybrid methods demonstrate comparable performance, with computation times remaining nearly identical, while the metaheuristic methods experience a higher rate of computation time growth. This suggests that, at smaller scales, the overhead introduced by hybridization does not yield significant gains over the classical baseline. However, as the problem size increases—particularly at 400 units and beyond—a clear divergence in performance emerges. The quantum–classical hybrid model consistently outperforms the classical MINLP, GA, and SA approaches, exhibiting noticeably lower computation times. The GA method had a mean computation time of 9,921 seconds at 200 units and the SA method had a mean computation time of 4573 seconds at 400 units, and both metaheuristic simulations were stopped there as they exceeded one hour of computation. This performance gap becomes increasingly pronounced with larger systems, underscoring the scalability advantage of the hybrid algorithm. For this study, computation time of each solver is compared, however the hybrid quantum-classical algorithm also has additional overhead time costs in the form of finding an embedding for the problem onto the QPU and latency in cloud communication [52]. However, the LeapHybridCQMSolver used for the CQM only reports time spent accessing the QPU, therefore the overhead time costs are not considered in this study. These overhead time costs can limit the utility of the hybrid model, particularly at smaller system sizes where the difference in computation time between MINLP and the hybrid algorithm can be relatively smaller. However, at larger scales, it is expected that the decrease in computation time from the hybrid model should eclipse the overhead costs, making them less relevant at larger, more realistic problem scales.

In its present state, QA provides similar solution times to MINLP, GA, and SA for small systems, but its performance curve scales more favorably with problem size, offering potential benefits as qubit counts and coherence times improve in future hardware generations. The advantage of leveraging D-Wave's CQM is that it reduces the burden of penalty coefficient tuning—traditionally a major challenge in QUBO modeling—and guarantees constraint satisfaction, thereby improving solution stability as system complexity grows.

Notably, at the 1,000-unit scale, the hybrid solution maintains computational feasibility where the classical method heuristic methods have already exceeded one hour of mean computation time. These results highlight the hybrid model's potential to efficiently handle large-scale resource scheduling problems, where classical methods often encounter computational bottlenecks. Figure 5 illustrates the mean computation time of the four methods using a logarithmic scale. From this, it can be seen that the computation time of the classical solutions increase at a much higher rate than the hybrid solution, highlighting the divergence in computation time between the UC solutions and the potential for the hybrid model to solve larger UC or resource scheduling problems more efficiently than current mixed-integer linear programming or classical metaheuristic methods. Results from a simplified discretized method have previously shown that the discretized QUBO model experiences approximately 59% growth in computation time from 10- to 200-unit systems [47], and results here for the classical model demonstrate a computation time growth of over 3,600% for the same range, while the decomposed hybrid model has experienced approximately a 61% computation time growth. Both the discretized and decomposed models exhibit a computation time growth far less than the classical-only model, and while the discretized model technically experienced less growth, it is representing a 1-hour system and its computation time can only be expected to grow further if expanded to a 24-hour system like the one presented here.

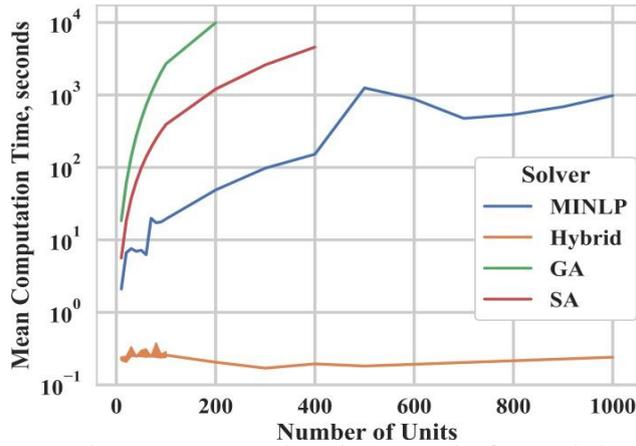

Figure 5: Computation Time Growth of UC Solutions

Due to the unit duplication method to create systems larger than 10 units, there exist multiple optimal solutions to larger systems, creating redundancies and symmetries in the solution space. Rather than compare multiple commitment schedules that have identical objective values but differing schedules, schedules are directly compared in terms of their objective values, by calculating the absolute cost error or the optimality gap between solutions as follows:

$$OG = \frac{|f_c - f_h|}{f_c} \tag{14}$$

where $OG$ is the overall optimality gap, $f_c$ is the objective value of the classical solution, and $f_h$ is the objective value of the hybrid quantum-classical solution. Figure 6 illustrates the optimality gap of the hybrid solution, or how far its objective is from the MINLP baseline solution.

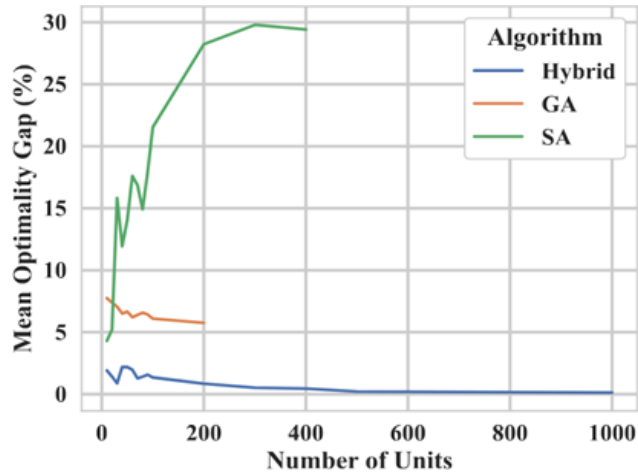

Figure 6: Optimality Gap Between UC Solutions

The optimality gap for the hybrid solution is highest at the lower system sizes where quantum discretization introduces the most approximation error, with an optimality gap of 2.19% at 40 units, and lowest at the 1000-unit system, with an optimality gap of 0.0133%. This indicates that the hybrid solution is never more than 2.19% less optimal than the classical solution, and that as the systems become larger and more complex, the hybrid solution is not experiencing a monotonically increasing optimality gap but rather that solution quality is increasing. The GA solution started at 7.75% optimality gap at 10 units and decreased to 5.76% at 200 units, while the SA solution started at 4.30% optimality gap at 10 units and increased to 29.42% at 400 units. The solutions for the metaheuristic methods are sensitive to their hyperparameters and would likely improve with more iterations, however their computation times quickly became too prohibitive to continue their simulations.

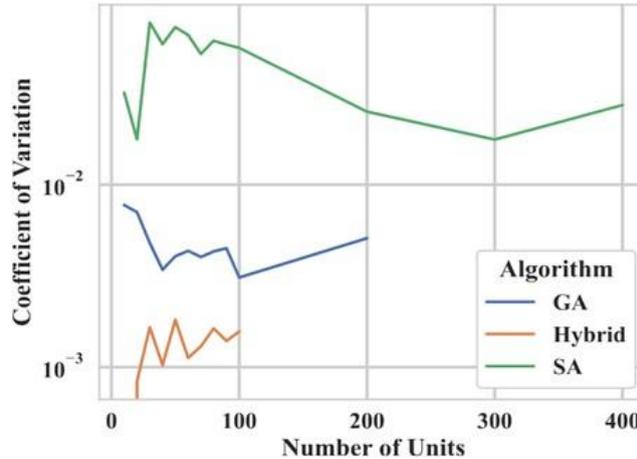

Figure 7: Coefficient of Variance of UC Solutions

To determine the variability of the proposed hybrid method compared to the classical methods, the coefficient of variation (CV) was calculated for each system size where multiple simulations were run. The CV is the ratio of the standard deviation of the objective values to the mean objective value for each system size. Figure 7 demonstrates the CV of each method across their simulation ranges. MINLP had a CV of 0 across all simulations, as it achieved the same result each time. SA consistently had the highest variation in results, followed by GA, then the hybrid algorithm. The variation of the hybrid algorithm's results increases as the system size increases, but it remains lower than the metaheuristic methods' variations. Alongside the computation time growth, these results indicate that the hybrid solution is well-suited for large-scale combinatorial optimization problems, like large-scale UC. In the range of 10- to 200-unit systems, the discretized model [47] experienced a monotonically increasing optimality gap, increasing from 3% to just over 20%. The decomposed hybrid model here experienced optimality gaps of 1.63% and 1.02% across the same range, indicating that the UC schedules it produces are consistently more optimal than those from the decomposed QUBO model. The lower optimality gap experienced by the proposed decomposition hybrid model could be due to the automatic handling of constraints by the CQM, rather than utilizing penalty coefficients in the QA-only QUBO model. The modeling of continuous power generation using a classical solver also likely improves the optimality gap of the proposed decomposed model, as more precise values can be optimized for, rather than approximations achieved by discretization.

Further investigations into the limits of hybrid modeling using decomposition methods for resource scheduling need to be investigated, as the results here demonstrate that a hybrid quantum-classical model can achieve near-optimal solutions, at most 1.63% from optimal, and demonstrate a computation time growth much lower than a classical-only model. However, the hybrid 24- hour UC model presented here is a simplified version to show the utility and feasibility of the hybrid quantum- classical algorithm, while UC used by grid operators involve aspects like security constraints, power flow constraints, and spinning reserves. These and other similar constraints will require additional qubits to be utilized to model them using existing QA architecture, and qubits are often a limiting factor in the scope of problems that can be embedded to the QPU and modeled successfully. As QPU architecture and embedding algorithms continue to develop and expand, it is expected that more complex, realistic UC and resource scheduling problems can be modeled. This work serves as an initial investigation into the utilization of QC and hybrid quantum-classical models for UC and resource scheduling. This study also assumes certain knowledge of demand profiles and generator parameters, while real-world power system operation often involves uncertainty in these inputs. Perturbations in demand or cost parameters would change the coefficients of the QUBO formulation, potentially affecting the balance of objective and constraint terms and, in turn, the resulting unit commitment schedule. Such sensitivity may require re-solving to maintain feasibility and near- optimality. Future work could address this challenge through scenario-based analysis, in which multiple QUBOs are generated and solved for perturbed demand scenarios to assess solution stability and performance variation. Using iterative penalty-tuning frameworks or the CQM framework, which automatically handles mathematical constraints, may improve solution robustness under input uncertainty. Incorporating these techniques would bring the proposed methodology closer to deployment in realistic, uncertain operating environments.

Additionally, as problem size and connectivity increase, quantum noise and decoherence are more likely to perturb the annealing process away from the ground state [53, 54]. These observations reinforce the need for advanced decomposition strategies, embedding heuristics, and error-mitigation techniques to scale hybrid quantum-classical UC to even larger, more realistic systems in future work.

## 5. Conclusions and Future Work

A quantum-classical hybrid optimization strategy utilizing Benders decomposition was proposed to improve the accuracy and computational efficiency of the unit commitment solution, with broader applicability to resource scheduling algorithms. The quantum component was implemented using D-Wave's Ocean software development kit, while the classical aspects were implemented using IBM's DOcplex. The hybrid model was compared to a classical MINLP model and metaheuristic GA and SA models across a range of 10- to 1,000-unit systems, and the results demonstrate that the hybrid model has a consistently lower computation time, lower rate of growth, and optimality gap within 2.19%.

The results illustrate the potential for hybrid computing to tackle the challenges introduced by the ever-increasing complexity of modern power grids. Quantum technology and solutions are developing rapidly, however pure quantum solutions are currently found to produce inaccurate results or be insufficient at decreasing computation time when it comes to the optimization of resource scheduling problems. Hybrid models may represent a method to bridge the gap between classical and quantum computers, utilizing the strength of both computational paradigms to provide adaptable solutions. This study provides an exploration and guide of hybrid computing, with initial results demonstrating a feasible, competitive solution to its classical counterparts.

This study intentionally utilized replicated 10-unit blocks to construct larger test systems, representing an initial step to developing and validating hybrid quantum-classical for the 24-hour UC problem and providing a controlled and scalable environment to evaluate algorithmic performance independently of system heterogeneity. This approach allows for isolating the effects of problem size on computational behavior and benchmarking the proposed hybrid quantum-classical framework at the upper bounds of current QPU capacity.

Future work will introduce random perturbations to reduce solution symmetry and incorporate operationally critical constraints—including transmission and security limits, spinning reserves, emissions limits, and power flow modeling—to bring the UC formulation closer to real-world practice. These extensions, along with continued advances in quantum hardware capacity and connectivity, will enable even more realistic and practically relevant hybrid UC solutions. We intend to investigate the application of hybrid quantum-classical modeling and Benders decomposition for other resource scheduling problems within the power systems domain as well. Further investigations into more complex, realistic modeling environments will require further advancements in quantum hardware and hybrid modeling methods.

Data and code for the simulations and analyses produced here are available upon request to the corresponding author.